\documentclass[journal]{IEEEtran}
\usepackage{mathrsfs}
\usepackage{bbm}
\usepackage{amssymb}
\usepackage{bbding}
\usepackage{threeparttable}

\usepackage[mathcal]{euscript}

\usepackage{psfrag,calc,url,bm}
\usepackage{diagbox}
\usepackage{cite}

\usepackage{graphicx}
\usepackage{siunitx}
\usepackage{psfrag}

\usepackage{subfigure}
\usepackage{hyperref}
\usepackage{url}

\usepackage{stfloats}

\usepackage{amsmath}

\usepackage{float}

\usepackage{algorithmic}

\usepackage[ruled,linesnumbered,vlined]{algorithm2e}

\usepackage{color}

\usepackage{boxedminipage}

\usepackage{amsthm}

\usepackage{multirow}

\usepackage{setspace}

\usepackage{soul}
\usepackage{epstopdf}

\usepackage{booktabs}

\usepackage{cases}
\allowdisplaybreaks[3]

\begin{document}

\title{SWIPTNet: A Unified Deep Learning Framework for SWIPT based on GNN and Transfer Learning}
\author{Hong Han, Yang Lu,~\IEEEmembership{Member,~IEEE}, Zihan Song, Ruichen Zhang, Wei Chen~\IEEEmembership{Senior Member,~IEEE},\\ Bo Ai,~\IEEEmembership{Fellow,~IEEE}, Dusit Niyato,~\IEEEmembership{Fellow,~IEEE}, and Dong In Kim,~\IEEEmembership{Fellow,~IEEE}
\thanks{Hong Han, Yang Lu and Zihan Song are with the School of Computer Science and Technology, Beijing Jiaotong University, Beijing 100044, China (e-mail: 24120330@bjtu.edu.cn, yanglu@bjtu.edu.cn, 23120405@bjtu.edu.cn).}
\thanks{Ruichen Zhang and Dusit Niyato are with the College of Computing and Data Science, Nanyang Technological University, Singapore 639798 (e-mail: ruichen.zhang@ntu.edu.sg, dniyato@ntu.edu.sg).}
\thanks{Wei Chen and Bo Ai are with the School of Electronics and Information Engineering, Beijing Jiaotong University, Beijing 100044, China (e-mail: weich@bjtu.edu.cn, boai@bjtu.edu.cn).}
\thanks{Dong In Kim is with the Department of Electrical and Computer Engineering, Sungkyunkwan University, Suwon 16419, South Korea (e-mail: dongin@skku.edu)}
}
\maketitle

\begin{abstract}
This paper investigates the deep learning  based approaches for simultaneous wireless information and power transfer (SWIPT). The quality-of-service (QoS) constrained sum-rate maximization problems are, respectively, formulated for power-splitting (PS) receivers and time-switching (TS) receivers and solved by a unified graph neural network (GNN) based model termed SWIPT net (SWIPTNet). To improve the performance of SWIPTNet, we first propose a single-type output method to reduce the learning complexity and facilitate the satisfaction of QoS constraints, and then, utilize the Laplace transform to enhance input features with the structural information. Besides, we adopt the multi-head attention and layer connection to enhance feature extracting. Furthermore, we present the implementation of transfer learning to the SWIPTNet between PS  and TS receivers. Ablation studies show the effectiveness of key components in the SWIPTNet. Numerical results also demonstrate the capability of SWIPTNet in achieving near-optimal performance with millisecond-level inference speed which is much faster than the traditional optimization algorithms. We also show the effectiveness of transfer learning via fast convergence and expressive capability improvement.
\end{abstract}

\begin{IEEEkeywords}
Deep learning, SWIPT, GNN, transfer learning.
\end{IEEEkeywords}

\section{Introduction}

\subsection{Background}

With the explosive growth of mobile devices, enhancing the wireless communication performance becomes an urgent need for large-scale wireless networks. Especially, in Internet of Things (IoT) involving huge number of low-power wireless devices, the wireless networks account for both information transfer and energy supply, which makes the simultaneous wireless information and power transfer (SWIPT) an important topic for future communication system\cite{bk1}. The power-splitting (PS) and time-switching (TS) receivers are two classic SWIPT receivers \cite{bk2}, which are, respectively, suitable for  high efficiency and easy implementation. Thanks to the deployment of the multiple antennas, the energy of radio frequency (RF) signals can be concentrated on the desired locations via beam-focusing, which greatly enhances the dual functions of SWIPT \cite{bk3}. Via frequency spectrum sharing, the trade-off between interference mitigation and utilization draws great attentions in SWIPT networks, which makes the transmit design the central role. In the past decade, the convex (CVX) optimization theory has been widely adopted to develop the transmit design for SWIPT networks \cite{bk5}. Although the (near-)optimality can be achieved by the CVX-based algorithms, the iterative framework hinders the implementation of these algorithms to realize the real-time signal processing for practical time-varying wireless scenarios.

Recently, applying deep learning (DL) to optimize wireless networks has emerged as a revolutionized technology to design resource allocation algorithms \cite{dl1}. The neural networks can be trained to learn and store the capability of solving the specific optimization problems via supervised \cite{dl2} or unsupervised learning \cite{dl3}. The well-trained neural networks have been validated to realize near-optimal and real-time resource allocation for several wireless networks. For instance, in \cite{MLP}, the fully connected multi-layer perceptrons (MLPs) were trained to approximate the non-linear mapping from the channel state information (CSI) to power allocation scheme to maximize the sum-rate in interference channel. In \cite{CNN}, the convolutional neural networks (CNNs) were utilized to approximate the iterative processes of weighted minimum mean square error (WMMSE) algorithm to address the maximization of system utility for multi-user multiple-input multiple-output (MU-MIMO) systems. Numerical results show that the two DL models are much faster than the WMMSE algorithm but with negligible performance loss. However, the MLPs and CNNs may be unable to scale to a larger problem size unseen during the training phase due to fixed input setting and suffer intolerable generalization loss in complicated wireless networks due to limited feature extracting capability. Unfortunately, the SWIPT networks involve multiple user equipments (UEs) frequently switching between active and inactive status and dual-functional services sharing the same resource budget. Therefore, developing powerful DL models to allocate resource becomes a crucial consideration for DL-enabled SWIPT networks.

\subsection{Related Works}

Thus far, the DL-enabled transmit design for wireless networks has been widely investigated in existing works. In addition to supervised learning in \cite{MLP} and \cite{CNN}, some works attempted to adopt unsupervised learning to alleviate the burden to acquire labeled training samples. In \cite{unsuper} and \cite{unsuper2}, two deep neural networks (DNNs) were, respectively, trained via unsupervised learning to optimize the beamforming design for large-scale antenna arrays and wireless networks with instantaneous and statistic constraints. However, traditional models, e.g., MLPs and CNNs, are not scalable to input, which challenges the implementation of these models in dynamically changing wireless networks. To exploit the graphical topology inherent in wireless networks, recent works have paid increasing attentions to graph neural networks (GNNs)\cite{lugnn}. In \cite{GNN2021}, the message-passing GNN (MPGNN) was introduced to  wireless resource management problem, which satisfies permutation equivariance property and  can generalize to large-scale problems. In \cite{GCN2023}, a unified framework to apply GNN to general design problems in wireless networks was given, followed by theoretical analysis of the effectiveness and efficiency of GNNs. In \cite{GNN2024}, a GNN-based robust downlink beamforming design was proposed for MU-MISO networks, and a model-based structure was adopted to simplify the beamforming vectors to key parameters. Nevertheless, these GNN-based models are based on vanilla message passing process, which may be hard to extract the hidden features. In \cite{GAT2024}, a graph attention network (GAT)-based algorithm was proposed for MU-MISO networks to solve sum-rate maximization problem subject to data rate requirement and total power budget. In \cite{icnet}, an edge enhanced GNN was design to solve the energy efficiency (EE) maximization problem for interference channel. In \cite{hg}, a heterogeneous graph learning was adopted to achieve physical-layer secure beamforming. The mentioned works demonstrate the superiority of applying DL in the optimization of wireless networks.

On the other hand, some works focused on the DL-enabled transmit design for SIWPT networks.  In \cite{DNNSWIPT}, a DNN-based model was proposed for TS receivers to infer the sum-rate and EE with low-computation complexity and high accuracy. In \cite{DNN2SWIPT}, the secrecy rate in single-antenna SWIPT maximized by a DNN-based model was shown to be very close to the successive CVX approximation method. In \cite{DNNS}, the DNN was combined with semidefinite relaxation technique to design a dual-layer algorithm to minimize the power consumption in SWIPT networks based on rate-splitting multiple access. In \cite{DNNSWIPT2}, a DNN-based model was trained for SWIPT to address the high computation time issue of the iterative algorithm via combining supervised and unsupervised training strategy. Nevertheless, \cite{DNNSWIPT,DNN2SWIPT,DNNS,DNNSWIPT2} paid little attention to architecture design of the neural networks, and few works attempted GNN in SWIPT networks. Moreover, the similarity between PS and TS receivers is neglected in existing works, which however, is the key point to facilitate the transfer learning \cite{TLwithLoss} in SWIPT networks.

\subsection{Contributions}

\emph{To the best of our knowledge, applying GNN to  solve the resource allocation problems for PS and TS receivers in SWIPT networks has not been studied in the existing works.} To fill this gap, we re-think the resource allocation problems in SWIPT networks from a perspective of DL, exploit the common knowledge domain for SWIPT networks via GNN, and transfer the knowledge between PS and TS receivers. The major contributions are summarized as follows:

\begin{itemize}
    \item We propose a unified DL framework for SWIPT networks, which is applicable to both PS and TS receivers. Via unsupervised learning formulation and sinlge-type output method, the resource allocation problems for PS and TS receivers are can be solved by one DL model trained on a sharing dataset. Particularly, we consider the classic QoS-constrained sum-rate maximization problems for MU-MISO SWIPT network based on PS and TS receivers, where the beamforming vectors and PS/TS ratios are required to be optimized.

    \item An efficient GNN-based model termed SWIPT net (SWIPTNet) is proposed to solve the considered problems by mapping the input CSI to desired transmission designs for PS and TS receivers, respectively. By modeling the SWIPT network as a graph, we consider the Laplace transform to enhance node features with graphical information. Then, we adopt the multi-head attention and layer connection , respectively, to enhance feature extracting and mitigate the over-smoothing issue due to stacking deep layers. The effectiveness of above mechanisms is analyzed, and the SWIPTNet is scalable to problem size due to parameter sharing.

    \item Considering the similarity in the knowledge domain of PS and TS receivers, we present the implementation of transfer learning to the SWIPTNet in order to improve the learning performance and reduce the training cost. Under the proposed DL framework, the knowledge can be transferred via the parameters of the well-trained model from the source task\footnote{In this paper, the task represents training a DL model to solve the considered problem.} to the target task, and both the hyperparameters of model and the training set are shared during the transfer learning.

\end{itemize}

We provide extensive experimental results to evaluate the proposed SWIPTNet as well as validating the proposed DL framework. The ablation experiment shows the effectiveness of Laplace transform, single-type output and layer connection to demonstrate their enhancement to the learning performance of SWIPTNet. The optimality performance, scalability performance, feasibility rate and inference time of the SWIPTNet are evaluated, and the results show that the SWIPTNet achieves near-optimal performance with millisecond-level inference speed which is much faster than the traditional optimization algorithms and outperforms existing models (i.e., MLP, graph convolutional networks (GCN) and GAT). Finally, the convergence behavior of  transfer learning is illustrated which is much faster than that of direct learning, and the SWIPTNet after a few epochs of transfer learning achieves comparative test performance with the direct learning.

The rest of this paper is organized as follows. Section II gives the system model and problem formulation. Section III presents the DL framework and SWIPTNet. Section IV extends the DL framework with the transfer learning. Section V provides numerical results. Finally, Section VI concludes the paper.

\emph{Notations}:
The following mathematical notations and symbols are used throughout this paper. $\bf a$ and $\bf A$ stand for a column vector and a matrix (including multiple-dimensional tensor), respectively. The sets of $n$-dimensional real column vector, $n$-by-$m$ real matrices and $n$-by-$m$-by-$k$ real tensors are denoted by ${\mathbb{R}^n}$, ${\mathbb{R}^{n \times m}}$ and ${\mathbb{R}^{n \times m \times k}}$, respectively. The sets of complex numbers, $n$-dimensional complex column vector and $n$-by-$m$ complex matrices are denoted by ${\mathbb{C}}$, ${\mathbb{C}^n}$ and ${\mathbb{C}^{n \times m}}$, respectively. For a complex number $a$, $\left| a \right|_2$ denotes its modulus and ${{\rm Real}(a)}$ denote its real part. For a vector $\bf a$, ${\left\| \bf a \right\|}$ is the Euclidean norm. For a matrix ${\bf A}$, ${\bf A}^H$ and $  \left \|{\bf A}\right\|$ denote its conjugate transpose and Frobenius norm, respectively. In addition, $[{\bf A}]_{i,j}$, $[{\bf A}]_{i,:}$ and $\left[{\bf A}\right]_{:,j}$ denote its $i$-th row and the $j$-th column element, the $i$-th row vector and the $j$-th column vector, respectively.

\section{System Model and Problem Formulation}

\begin{figure}
\centering
\includegraphics[width=0.49\textwidth]{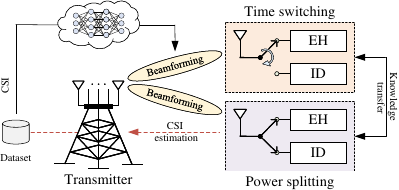}
 \caption{Illustration of the unified DL framework for SWIPT networks where the CSI acquired from the UEs is mapping to the desired beamforming vectors and the knowledge is transferred between the TS and PS receivers. }\label{sys}
\end{figure}

Consider a downlink SWIPT network as shown in Fig. \ref{sys}, where a $N_{\rm T}$-antenna transmitter serves $N$ energy-constrained UEs.  For clarity, we use $n$ to denote the $n$-th UE, where $n \in \mathcal{N} \buildrel \Delta \over =  \left\{1,2, ..., N\right\}$.

In each time slot, the transmit signal is given by
\begin{flalign}
{\bf x} =\sum\nolimits_{n = 1}^N { {\bf w}_n {s _n}},
\end{flalign}
where ${{{\bf w}}_n} \in {\mathbb{C}^{{N_{\rm{T}}} }}$ and $s_n \in \mathbb{C}$ denote the beamforming vector and symbol associated with the $n$-th UE, respectively. Without loss of generality, we consider that ${{\mathbb E}}\{ {{{| {s_n } |}^2}} \} = 1$,

Denote ${\bf h}_n \in \mathbb{C}^{N_{\rm T}}$ to be the CSI\footnote{The acquisition of (perfect) CSI is  a critical factor for the low-power IoT applications. Some channel tracking schemes based on DL can be adopted like \cite{CSI}, and the DL-based CSI estimation module can be cascaded with the DL-based resource allocation module as the output of the former is the input of the latter.} vectors from the transmitter to the $n$-th UE. For the $n$-th UE, the received signal is given by
\begin{flalign}
{{y}}_n= {\bf h}^H_n{\bf x} &=  {{\bf{h}}^H_n{{\bf{w}}_n}{s _n}} + {\sum\nolimits_{m \ne n}^N {{\bf{h}}^H_n{{\bf{w}}_m}{s _m}} } + {{{n}}_n},\label{rec_IR}
\end{flalign}
where ${ n}_n\sim\mathcal{CN}\left( {0,{\sigma_{n}^2}}\right)$ denotes the additive white Gaussian noise (AWGN) at the $n$-th UE with ${\sigma_{n}^2}$ denoting the noise power. Since $\sigma_{n}^2$ is generally much smaller than the noise power $\sigma_{\rm s}^2$ introduced by the baseband processing circuit, $\sigma_{n}^2$ is neglected in the following analysis.

Typically, there are two types of SWIPT receivers \footnote{Although the TS receiver can be regarded as a special case of the PS receiver mathematically, it has an advantage in low hardware complexity and power consumption in view of practical implementation. Therefore, we consider the two SWIPT receivers, which however, can be optimized by a unified DL model.}, i.e.,
PS receiver and TS receiver. Accordingly, the sum-rate maximization problems under the two SWIPT receivers are, respectively, formulated in the subsequent two subsections.

\subsection{Power-Splitting Receiver}

With PS receiver, the received signals at the $n$-th UE are split into two streams, where $\sqrt \rho_n$ part is used for information decoding (ID) and the rest $\sqrt{1-\rho_n}$ part is used for energy harvesting (EH). Following (\ref{rec_IR}), the achievable information rate of the $n$-th UE is given by
\begin{flalign}\label{rps}
{R^{\rm PS}_n}&\left( {\left\{{\bf{w}}_i\right\},\rho_n } \right) =\log \left( {1 + \frac{{\rho_n}{{{\left| {{\bf{h}}_n^H{{\bf{w}}_n}} \right|}^2}}}{{\rho_n}{\left(\sum\nolimits_{m \ne n}^N {{{\left| {{\bf{h}}_n^H{{\bf{w}}_m}} \right|}^2}}\right)  + {\sigma_{\rm s} ^2}}}} \right).
\end{flalign}

At the $n$-th UE, $\sqrt{1-\rho_n}$ part of received signals is for EH, i.e.,
\begin{flalign}
{\gamma^{\rm PS}_n}\left( {\left\{{\bf{w}}_i\right\},\rho_n } \right) = \left( {1 - {\rho _n}} \right)\sum\nolimits_{i = 1}^N {{{\left| {{\bf{h}}_n^H{{\bf{w}}_i}} \right|}^2}}.
\end{flalign}

Our goal is to maximize the sum achievable rates of the considered system subject to the quality-of-service (QoS) requirements of UEs, which is formulated as follows:
\begin{subequations}\label{p1}
\begin{align}
&\mathop {\max }\limits_{ {\left\{ {{{\bf{w}}_{i}},\rho_i} \right\}}}\sum\nolimits_{n = 1}^N {{R^{\rm PS}_n}} \left( {\left\{ {{{\bf{w}}_i}} \right\},{\rho _n}} \right) \\
{\rm s.t.}~&{R^{\rm PS}_n}\left( {\left\{{\bf{w}}_i\right\},\rho_n } \right)  \ge {R}_{\rm req},\label{primal:A}\\
&{\gamma^{\rm PS}_n}\left( {\left\{{\bf{w}}_i\right\},\rho_n } \right) \ge {{\gamma}_{\rm req}},\label{primal:B}\\
&\sum\nolimits_{n = 1}^N {\left\| {{{\bf{w}}_n}} \right\|_2^2} \le P_{\rm max},\label{primal:C}\\
&0\le \rho_n \le 1, {\forall n \in \cal N},
\end{align}
\end{subequations}
where ${R}_{\rm req}$ and ${{\gamma}_{\rm req}}$ respectively denote the UE's individual rate requirement and EH requirement\footnote{The EH requirement ${{\gamma}_{\rm req}}$ is normalized by the non-linear EH model \cite{non-linear}.}, and $P_{\rm max}$ denotes the power budget of the transmitter.

\subsection{Time-Switching Receiver}

With time-switching receiver, each transmission block of the $n$-th UE is divided into two orthogonal time slots with $\alpha_n$ percentage for ID and $(1-\alpha_n)$ percentage for EH.

Following (\ref{rec_IR}), the achievable information rate of the $n$-th UE is given by
\begin{flalign}\label{p6}
{R^{\rm TS}_n}\left( {\left\{{\bf{w}}_i\right\}} \right) =\log \left( {1 + \frac{{{{\left| {{\bf{h}}_n^H{{\bf{w}}_n}} \right|}^2}}}{{\sum\nolimits_{m \ne n}^N {{{\left| {{\bf{h}}_n^H{{\bf{w}}_m}} \right|}^2}}  + {\sigma_{\rm s} ^2}}}} \right).
\end{flalign}

The power received at the $n$-th UE is given by \begin{flalign}\label{p7}
{\gamma^{\rm TS}_n}\left( {\left\{{\bf{w}}_i\right\}} \right) = \sum\nolimits_{i = 1}^N {{{\left| {{\bf{h}}_n^H{{\bf{w}}_i}} \right|}^2}}.
\end{flalign}

Then, the QoS-constrained sum-rate maximization problem is formulated as follows:
\begin{subequations}\label{p2}
\begin{align}
&\mathop {\max }\limits_{ {\left\{ {{{\bf{w}}_{i}},\alpha_i} \right\}}}\sum\nolimits_{n = 1}^N \alpha_n {{R^{\rm TS}_n}} \left( {\left\{ {{{\bf{w}}_i}} \right\}} \right) \\
{\rm s.t.}~&{R^{\rm TS}_n}\left( {\left\{{\bf{w}}_i\right\} } \right)  \ge {R}_{\rm req}/\alpha_n,\label{p2:A}\\
&{\gamma^{\rm TS}_n}\left( {\left\{{\bf{w}}_i\right\}} \right) \ge {{\gamma}_{\rm req}}/\left(1-\alpha_n\right),\label{p2:B}\\
&\sum\nolimits_{n = 1}^N {\left\| {{{\bf{w}}_n}} \right\|_2^2} \le P_{\rm max},\label{p2:C}\\
&0\le \alpha_n \le 1, {\forall n \in \cal N}.\label{p2:D}
\end{align}
\end{subequations}

Both problems (\ref{p1}) and (\ref{p2}) are classical resource allocation problems for SWIPT network and can be solved by the traditional CVX algorithms which however, is hard to match the time-varying wireless environment due to time-costly iterations. It is interesting to see that the variables in problem (\ref{p1}), i.e., ${\left\{ {{{\bf{w}}_{i}},\rho_i} \right\}}$, and the ones in problem (\ref{p2}), i.e., ${\left\{ {{{\bf{w}}_{i}},\alpha_i} \right\}}$, have the same shape and the same feasible set. Such an observation motivates us to design a unified DL framework to solve the problems (\ref{p1}) and (\ref{p2}).

\section{DL framework and SWIPTNet}

\begin{table*}[ht]\label{table:pare}
	\caption{Summary of Main Notations Used in Architecture of SWIPTNet}
	\centering
	\begin{tabular}{c|c}
		\hline
		{\bf Notation} & {\bf Definition} \\
		\hline\hline
	      ${\bf h}_n \in \mathbb{C}^{N_{\rm T}}$ & Input feature of the $n$-th node. \\
        \hline
            ${\bf U} \in {\mathbb R}^{N\times N}$ &   The matrix of eigenvectors of the Laplace matrix of ${\cal G}$.   \\
        \hline
            ${\widehat{{\bf h}}_n} \in \mathbb{C}^{N_{\rm T} + N}$ &  The $n$-th enhanced node features with Laplace transform. \\
        \hline
            $\overline{\bf h}_n \in {\mathbb C}^{\overline l}$ & Input feature of the $n$-th node of the GAL with the input dimension of ${\overline l}$. \\
        \hline
            $\widetilde{\bf h}_n\in {\mathbb C}^{\widetilde l}$ & Output feature of the $n$-th node of the GAL with the output dimension of ${\widetilde l}$. \\
        \hline
            $\alpha_{n\leftarrow j}^{\left(s\right)} \in {\mathbb R}$ & The $s$-th attention coefficient between the $n$-th node and the $j$-th node.\\
        \hline
             $\mathrm{{\bf a}}^{(s)} \in {\mathbb C}^{\widetilde l} $, ${\bf W}^{(s)}_{\rm dir}, {\bf W}^{(s)}_{\rm ner} \in {\mathbb C}^{\widetilde l \times {\overline l}}$ & Learnable parameters of $s$-th attention head of the GAL.\\
        \hline
            $\boldsymbol{\beta}_n^{(s)}$ &
            The received message of $s$-th attention head of the $n$-th node.\\
        \hline
            $\overline{\bf y}_n \in {\mathbb C}^{\overline d}$ & Input feature of the $n$-th node of the FL with the input dimension of ${\overline d}$.\\
        \hline
            $\widetilde{\bf y}_n\in {\mathbb C}^{\widetilde d}$ & Output feature of the $n$-th node of the FL with the input dimension of ${\widetilde d}$.\\
        \hline
            ${\bf Q} \in {\mathbb C}^{\widetilde d \times \overline d}$ & Learnable parameters of the FL.\\
        \hline
	\end{tabular}
\end{table*}

The main idea of the proposed DL approach is given here. First, we reformulate the problems (\ref{p1}) and (\ref{p2}) to facilitate the unsupervised learning where the constraints can be satisfied by either scale operation or penalty method. Then, the reformulated problem is handled by the SWIPTNet to realize the end-to-end learning from CSI vectors to beamforming vectors. At last, we use a PS/TS ratio recover module to obtain desired transmit design i.e., $\{ {\bf w}_i, \rho_i \}$ and $\{ {\bf w}_i, \alpha_i \}$.

\subsection{Unsupervised Learning Formulation}

To reduce the burden on collecting labeled training dataset, the unsupervised learning is adopted. In order to obtain feasible solutions, the constraints in problem (\ref{p1}) and problem (\ref{p2}) are handled as follows.

\subsubsection{Sum-Power Budget Constraint}
To guarantee that the output $\{{{\bf{w}}_{i}}\}$ meets the power budget constraint, i.e. \eqref{primal:C} and \eqref{p2:C}, the following scale operation is defined as follows:
\begin{flalign}\label{af}
\Phi \left( {{\bf{w}}}_n \right)
= \sqrt {\frac{{{P_{\rm max }}}} {{\rm max}\left({{P_{\rm max }}},\sum\nolimits_{i \in {\cal N} } {{\left\| {{{\bf{w}}_i}} \right\|}^2}\right)}}{{\bf{w}}_n}.
\end{flalign}

\subsubsection{PS Ratio and QoS Constraints in Problem (\ref{p1})}

For problem \eqref{p1}, with given $\{{{\bf{w}}_{i}}\}$, the EH constraint, i.e., \eqref{primal:B}, is guaranteed to be satisfied by setting the PS ratio to be
\begin{flalign}\label{dlrho}
{\rho _n} = 1 -  \frac{{\gamma}_{\rm req}}{\sum\nolimits_{i = 1}^N {{{\left| {{\bf{h}}_n^H{{\bf{w}}_i}} \right|}^2}}},~\forall n.
\end{flalign}
To guarantee that $\rho_n$ in \eqref{dlrho} satisfies that $\rho_n\in \left[0,1\right]$, the following constraint is required, i.e.,
\begin{flalign}\label{cons:gamma}
{{\gamma}_{\rm req}} \le  {\sum\nolimits_{i = 1}^N {{{\left| {{\bf{h}}_n^H{{\bf{w}}_i}} \right|}^2}}},
\end{flalign}
such that only one type, i.e., beamforming vector, is required to output by the neural networks, i.e., \emph{single-type output}.

Then, by substituting \eqref{dlrho} into $R^{\rm PS}_n( {\{{\bf{w}}_i\},\rho_n } )$ in \eqref{rps}, we have that
\begin{flalign}
&{{\widehat R}^{\rm PS}_n}\left( {\left\{{\bf{w}}_i\right\} } \right) = \\
&\log \left( {1 + \frac{\left(1 -  \frac{{\gamma}_{\rm req}}{\sum\nolimits_{i = 1}^N {{{\left| {{\bf{h}}_n^H{{\bf{w}}_i}} \right|}^2}}}\right){{{\left| {{\bf{h}}_n^H{{\bf{w}}_n}} \right|}^2}}}{\left(1 -  \frac{{\gamma}_{\rm req}}{\sum\nolimits_{i = 1}^N {{{\left| {{\bf{h}}_n^H{{\bf{w}}_i}} \right|}^2}}}\right){\left(\sum\nolimits_{m \ne n}^N {{{\left| {{\bf{h}}_n^H{{\bf{w}}_m}} \right|}^2}}\right)  + {\sigma_{\rm s} ^2}}}} \right). \nonumber
\end{flalign}

The penalty method is adopted to meet the constraints in \eqref{primal:A} and \eqref{cons:gamma}. Then, the multi-objective loss function for problem \eqref{p1} is give by
\begin{flalign}\label{l1}
{\cal L}^{\rm PS}\left(\{{{{\bf w}_i}} \}\right) =
&- \lambda^{\rm PS}_0 \sum\nolimits_{n = 1}^N {{{\widehat R}^{\rm PS}_n}} \left( {\left\{ {{{\bf{w}}_i}} \right\}} \right) +\\
& \lambda^{\rm PS}_1 \sum\nolimits_{n = 1}^N {\rm ReLU} \left({R}_{\rm req} - {{{\widehat R}^{\rm PS}_n}} \left( {\left\{ {{{\bf{w}}_i}} \right\}} \right) \right) +\nonumber \\
& \lambda^{\rm PS}_2 \sum\nolimits_{n = 1}^N {\rm ReLU} \left({{\gamma}_{\rm req}} -  {\sum\nolimits_{i = 1}^N {{{\left| {{\bf{h}}_n^H{{\bf{w}}_i}} \right|}^2}}} \right), \nonumber
\end{flalign}
where $\lambda_0^{\rm PS}$,  $\lambda_1^{\rm PS}$ and  $\lambda_2^{\rm PS}$ are hyperparameters for multi-objective learning for the PS receiver.

\newtheorem{Remark}{Remark}

\begin{Remark}\label{Rem0}
(Effectiveness of Single-type output) The single-type output method helps to enhance feasibility at the cost of limited performance loss. By representing ${\rho_n}$ by $\{ {\bf w}_i\}$ via \eqref{dlrho}, the output of the neural networks is simplified which helps to improve the learning performance. Note that the EH constraints, i.e., \eqref{primal:B}, may not always hold equivalent to the optimal solution to the problem (\ref{p1}), which indicates that \eqref{dlrho} may induce performance loss. Nevertheless, the harvested energy of each UE intends to be equal to its EH requirement such that more resource can be used for data transmission. Therefore, the optimality loss due to \eqref{dlrho} is limited.
\end{Remark}

\subsubsection{TS Ratio and QoS Constraints in Problem (\ref{p2})}

For the problem \eqref{p2}, we set
\begin{flalign}\label{dlalpha}
\alpha_i =1- \frac{{{\gamma}_{\rm req}}}{\sum\nolimits_{i = 1}^N {{{\left| {{\bf{h}}_n^H{{\bf{w}}_i}} \right|}^2}}},~\forall n,
\end{flalign}
following the single-type output method to guarantee that the EH constraint can be satisfied. With \eqref{cons:gamma}, $\alpha_i$ in \eqref{dlalpha} is guaranteed to satisfy
$\alpha_i\in \left[0,1\right]$.

The penalty method is adopted to make \eqref{p2:D} and \eqref{dlalpha} to be satisfied. Then, the loss function for problem (\ref{p2}) is give by
\begin{flalign}\label{l2}
{\cal L}^{\rm TS}&\left(\{{{{\bf w}_i}} \}\right) =
- \lambda^{\rm TS}_0 \sum\nolimits_{n = 1}^N \alpha_n {{R^{\rm TS}_n}} \left( {\left\{ {{{\bf{w}}_i}} \right\}} \right) + \\
& \lambda^{\rm TS}_1 \sum\nolimits_{n = 1}^N {\rm ReLU} \left({R}_{\rm req}/\alpha_n - {{{ R}^{\rm TS}_n}} \left( {\left\{ {{{\bf{w}}_i}} \right\}} \right) \right) +\nonumber \\
& \lambda^{\rm TS}_2 \sum\nolimits_{n = 1}^N {\rm ReLU} \left({{\gamma}_{\rm req}}/\left(1-\alpha_n\right) -  {\gamma^{\rm TS}_n}\left( {\left\{{\bf{w}}_i\right\}} \right) \right), \nonumber
\end{flalign}
where $\lambda_0^{\rm TS}$,  $\lambda_1^{\rm TS}$ and  $\lambda_2^{\rm TS}$ are hyperparameters for multi-objective learning for the TS receiver.

\subsection{Architecture of SWIPTNet}

\begin{figure*}
\centering
\includegraphics[width=1\textwidth]{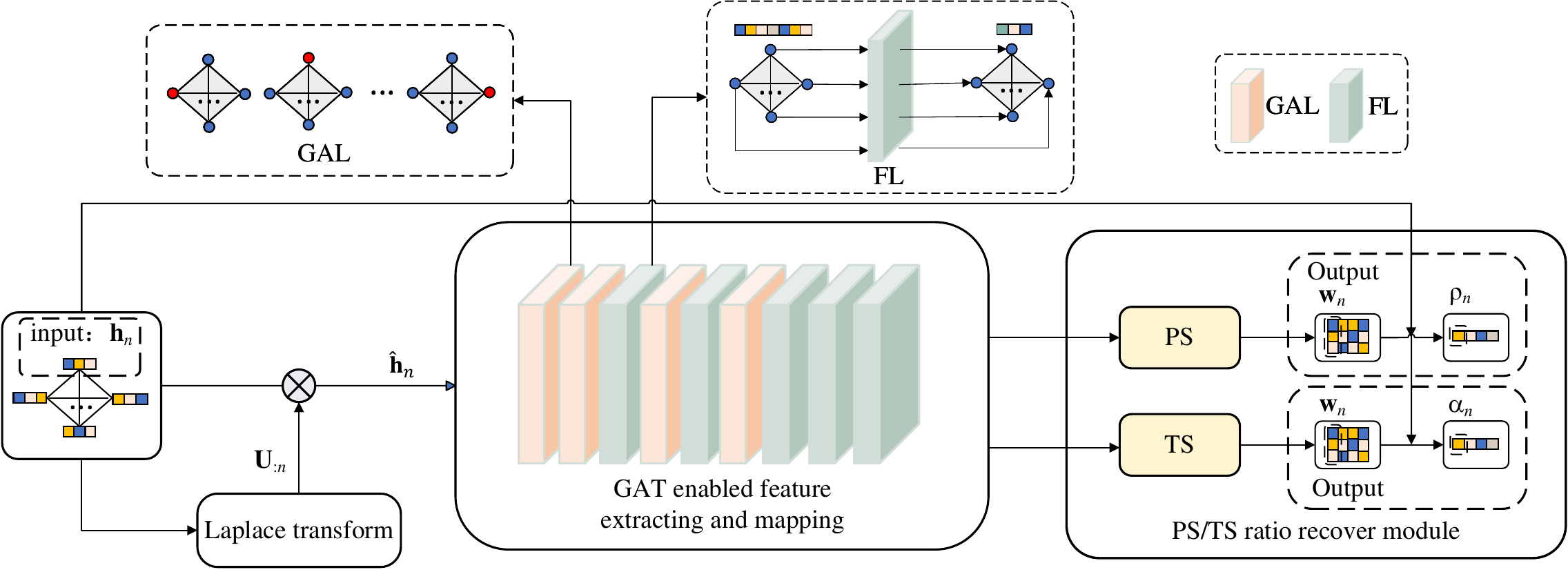}
 \caption{The structure of the SWIPTNet.}\label{GAT}
\end{figure*}

After the above reformulation, the neural networks are required to map CSI ${\left\{ {{{\bf{h}}_{i}}} \right\}}$ to  ${\left\{ {{{\bf{w}}_{i}}} \right\}}$. As shown in Fig. \ref{GAT}, the SWIPTNet to realize the mapping consists of three ingredients, i.e., 1) graphical representation, 2) Laplace transform based feature enhancement, 3) GNN enabled feature extracting and mapping. The detailed processes of each ingredient are introduced as follows.

\subsubsection{Graphical Representation}
\begin{figure}
\centering
\includegraphics[width=0.40\textwidth]{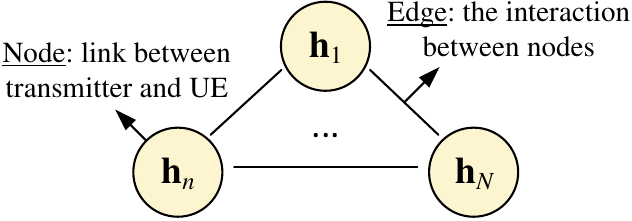}
 \caption{Graphical representation of the SWIPT network.}\label{node}
\end{figure}

As shown in Fig. \ref{node}, the SWIPT network can be represented by a fully-connected undirected graph, i.e., ${\cal G}=({\cal V},{\cal E})$, where ${\cal V}$ denotes the set of nodes with $\|{\cal V}\|=N$ and ${\cal E}$ denotes the set of edges with $\|{\cal E}\| = N(N-1)$. The $i$-th node represents the link between the transmitter and the $i$-th UE with the feature of ${\bf h}_i$. The edge between the $i$-th node and the $j$-th node represents the existence of the interaction between the two nodes with no feature. With the defined ${\cal G}$, the adjacency matrix is given by
\begin{flalign}
\left[{\bf A}\right]_{i,j} = \left\{ \begin{array}{l}
0,~i = j\\
1,~i \ne j
\end{array} \right.,
\end{flalign}
and the set of the neighboring nodes of the $i$-th node is defined by ${\widehat {\cal N}}(i)={\cal N}\setminus \{i\}$.

\subsubsection{Laplace Transform Based Feature Enhancement}

It is observed that the node features, i.e., $\{{\bf h}_{i}\}$, do not involve the interactions among links. Therefore, we enhance the node features based on the Laplace matrix of ${\cal G}$ which is defined by ${\bf L} \in {\mathbb R}^{N\times N}$ and given by
\begin{flalign}\label{eq:laplace}
[{\bf L}]_{i,j} = [{\bf D}]_{i,j} - [{\bf A}]_{i,j} = \left\{ \begin{array}{l}
[{\bf D}]_{i,i},  i=j\\
-1, i \ne j
\end{array} \right.,
\end{flalign}
where ${\bf D} \in {\mathbb R}^{N\times N}$ denotes the degree matrix of $\cal G$ with $[{\bf D}]_{i,i}= \sum\nolimits_{j = 1}^N {[{\bf A}]_{i,j}}$ = $N-1$ ($\forall i$).

The definition in \eqref{eq:laplace} shows that $ {\bf L} $ is full-rank. Then, ${\bf L}$ can be written as an eigenvalue decomposition, i.e.,
\begin{flalign}
    {\bf L} = {\bf U} {\bm \Lambda} {\bf U}^{T},
\end{flalign}
where  ${\bf U} \in {\mathbb R}^{N\times N} $ denotes the matrix of eigenvectors of $ {\bf L}$ and ${\bm \Lambda} \in {\mathbb R}^{N\times N}$ is a diagonal matrix with the corresponding eigenvalues on the diagonal.

Denote $\{\widehat{{\bf h}}_i\}$ by the enhanced node features, which is given by
\begin{flalign}\label{eh_f}
\widehat{{\bf h}}_i = {\rm Concat}\left({{\bf h}}_i, \left[{\bf U}\right]_{:i}+{\rm i}\left[{\bf U}\right]_{:i}  \right) \in {\mathbb C}^{(N_{\rm T}+N)},
\end{flalign}
where ${\rm Concat}(\cdot)$ represents the concatenation operation.

\begin{Remark}\label{Rem2} (Effectiveness of Laplace transform) The Laplace transform enhances the vanilla features with the structural information of the graph, which is also the main goal of the subsequent message passing process. It is observed for the $n$-th node, its vanilla feature, i.e., ${\bf h}_n$, does not involve any structural information about other nodes. After the  Laplace transform, the structural information helps each link to notice the existing of and the impact to/from other links and thus, prevents high inter-UE interference to strain information rates of some UEs with poor channel conditions. Besides, another advantage of the Laplace transform is parameter-free, which has no impact on the inference speed.
\end{Remark}

\subsubsection{GAT-Enabled Feature Extracting and Mapping}

The enhanced node features are input to the SWIPTNet which consists of two types of layers, i.e., the graph attention layer (GAL) and fully-connected layer (FL). The GAL is to extract the hidden features among nodes while the FL is to map the extracted feature to desired dimensions. As shown in Fig. \ref{GAT}, the two types of layers are stacked alternately to enhance the learning performance. The detailed processes of the two layers are given as follows.

\paragraph{GAL} Denote the input\footnote{For the first GAL, $\overline{\bf h}_n=\widehat{\bf h}_n$ ($\forall n$) and ${\overline l} = N_{\rm T} +N$.} and output of a GAL by $\overline{\bf h}_n \in {\mathbb C}^{\overline l}$ and $\widetilde{\bf h}_n\in {\mathbb C}^{\widetilde l}$, respectively, where ${\overline l}$ and ${\widetilde l}$ represent the corresponding dimensions. The GAL involves two key computations, i.e., multi-head attention and message passing, where the multi-head attention is to help each node to distinguish the impacts from its neighboring nodes and the message passing is to extract hidden interaction among nodes to update the node features.

Denote $S$ as the number of attention heads. Then, the $s$-th ($s \in \{1,2, ..., S\}$) attention coefficient between the $n$-th node and the $j$-th node is given by
\begin{flalign}
    &\alpha_{n\leftarrow j}^{\left(s\right)}= \\
    & \frac{\exp \left(f_{2}\left(f_{1}\left( {\bf W}^{(s)}_{\rm dir} \overline{{\bf h}}_{n} + {\bf W}^{(s)}_{\rm ner} \overline{{\bf h}}_{j} \right)^{T} \mathrm{{\bf a}}^{(s)} \right)\right) }{\sum_{k \in {\widehat {\cal N}}(n)} \exp \left(f_{2}\left(f_{1}\left({\bf W}^{(s)}_{\rm dir} \overline{{\bf h}}_{n} + {\bf W}^{(s)}_{\rm ner} \overline{{\bf h}}_{k}\right)^{T}\mathrm{{\bf a}}^{(s)}\right)\right)} ,\nonumber
\end{flalign}
where $\mathrm{{\bf a}}^{(s)} \in {\mathbb C}^{\widetilde l}$, ${\bf W}^{(s)}_{\rm dir} \in {\mathbb C}^{\widetilde l \times {\overline l}}$ and ${\bf W}^{(s)}_{\rm ner} \in {\mathbb C}^{\widetilde l \times {\overline l}}$ denote learnable parameters corresponding to the $s$-th attention, $f_{1}(\cdot)$ represents a non-linear function, (e.g., $\operatorname{LeakyReLu}(\cdot)$), and $f_{2}(\cdot):{\mathbb C}\rightarrow{\mathbb R}$ represents a complex-to-real function (e.g., $\operatorname{Real}(\cdot)$).

With the obtained attention coefficients, the message passing enables each node to  first aggregate the weighted features from its neighboring nodes and then update its node features with all aggregated features. For the $n$-th node, the $s$-th attention enabled  feature aggregation is given by
\begin{flalign}
    \boldsymbol{\beta}
    _n^{(s)} = f_{3}\left( \alpha_{n\leftarrow  j}^{(s)} {\bf W}^{(s)}_{\rm ner} \overline{\bf h}_j \mid j \in {\widehat {\cal N}}(n) \right) \in {\mathbb C}^{\widetilde{l}} ,
\end{flalign}
where $f_{3}(\cdot):{\mathbb C}^{\widetilde{l} \times (N - 1)}\rightarrow{\mathbb C}^{\widetilde{l}}$ represents a function with permutation invariance (e.g., $\operatorname{Sum}(\cdot)$). By concatenating $S$ aggregated features, the updated feature, i.e, the output of GAL, is given by
\begin{flalign}
    \widetilde{\bf h}_n= {\rm AC}\left({\rm Concat}\left(\left\{\boldsymbol{\beta}_n^{(s)}\right\}_{s=1}^{S}\right)\right) \in {\mathbb C}^{S \times \widetilde{l}},
\end{flalign}
where ${\rm AC}(\cdot)$ represents the activation function (e.g. ${\rm SELU}(\cdot)$).

\paragraph{FL}

Denote the input and output\footnote{For the last FL, $\widetilde{\bf y}_n={\bf w}_n$ and ${\widetilde d}= N_{\rm T}$.} of an FL by $\overline{\bf y}_n \in {\mathbb C}^{\overline d}$ and $\widetilde{\bf y}_n\in {\mathbb C}^{\widetilde d}$, respectively, where ${\overline d}$ and ${\widetilde d}$ represent the corresponding dimensions. For each FL, the mapping from the input to the output is computed by
\begin{flalign}\label{FL}
    \widetilde{\bf y}_n = {\rm AC}\Big({\rm BN}\Big( {\underbrace {{\bf{Q}}{{\overline {\bf{y}} }_n}}_{ \buildrel \Delta \over = {{\dot{\bf{y}}}_n}}}\Big)\Big),
\end{flalign}
where ${\bf Q} \in {\mathbb C}^{\widetilde d \times \overline d}$ denotes learnable parameters of FL, and ${\rm BN}(\cdot)$ represents the batch normalization operation to normalize the FL output to reduce  internal covariate shift and speed up the model training process.

\emph{Note that the FL is utilized for two purposes, i.e., layer connection \cite{over-smoothing} when it is stacked between two GALs and transmit design decoder when it is stacked after the last GAL.} Only the FL for decoder purpose requires the BN operation. The FL for layer connection helps to stack deeper GNN to enhance the feature extracting capability.

\begin{Remark} \label{remark3} (Effectiveness of layer connection) The FL for layer connection is to mitigate the over-smoothing issue, since the node features become indistinguishable, which fails the training of GNN-based model, after several message-passing iterations. By adding the FL between GALs, the difference among the output node features of the former GAL can be enlarged.
\end{Remark}

In a summary, the learnable parameters in the SWIPTNet is represented by $\{\mathrm{{\bf a}}^{(s)}, {\bf W}^{(s)}_{\rm dir}, {\bf W}^{(s)}_{\rm ner}, {\bf Q}\}$. It is observed that the dimension of learnable parameters is independent of $N$. The reason is that $N$ nodes share learnable parameters in both GAL and FL. Such a design is able to facilitate the SWIPTNet to be scalable to $N$, i.e., the number of UEs.

\subsection{PS/TS Ratio Recovery Module}

The above processes (in Section III-A and Section III-B) map a given CSI realization, i.e., $\{ {\bf h}_i \}$, to a beamforming design, i.e., $\{ {\bf w}_i \}$. Therefore, the PS/TS ratio recovery module is to obtain $\{\rho_i\}$ or $\{\alpha_i\}$. For the PS/TS receiver, $\{\rho_i\}$/$\{\alpha_i\}$ are obtained via \eqref{dlrho}/\eqref{dlalpha}.

Note that the PS/TS ratio recovery module does not work during the training and test phases via the unsupervised learning formulation, the loss function \eqref{l1}/\eqref{l2} does not involves \eqref{dlrho}/\eqref{dlalpha}. However, the PS/TS ratio recovery module is required to be activated for the implementation of the model.

\section{Transfer learning}

It is observed that via the unsupervised learning formulation in Section III-A, the PS receiver and the TS receiver can share the same training set and model but use different loss functions. Therefore, one can fine-tune the well-trained model for PS/TS receiver to adapt to TS/PS receiver via transfer learning at a marginal training cost. In this section, we briefly describe the background of transfer learning and then give the detailed implementation of transfer learning to SWIPTNet.

\subsection{Preliminary Knowledge of Transfer Learning}

The transfer learning is to utilize the knowledge learned in one domain/task to other domains/tasks to achieve faster or better inference. Typically, we give the definitions of domain and task following by the concept of transfer learning, which is the fundamental that why the transfer learning can be used in the considered problem.

\subsubsection{Domain} {A domain $\cal{D}$} consists of two components, i.e., a $G$-dimension feature space denoted by $\mathbb{G}^G$ and the corresponding marginal probability distribution $P({\bf g})$ with ${\bf g}\in\mathbb{G}^G$.

\subsubsection{Task} Given a specific domain $\cal{D}$, a task  $\cal{T}$ consists of two components, i.e., a $T$-dimension label space denoted by $\mathbb{Y}^T$ and an objective mapping function $\Pi_{\bm \theta}(\cdot):\mathbb{G}^G\rightarrow\mathbb{Y}^T$, where ${\bm \theta}$ denotes the learnable parameters learned from the training data. Note that, in this paper, the label space represents the space of the objective function results, as the unsupervised learning and optimization problem are considered.

\subsubsection{Transfer learning} Given a source domain $\cal{D}_{\rm S}$ associated with a source learning task $\cal{T}_{\rm S}$ and a target domain $\cal{D}_{\rm T}$ associated with a target learning task $\cal{T}_{\rm T}$, the transfer learning aims to improve the learning of the target objective mapping function $\Pi_{{\bm \theta}_{\rm T}}(\cdot)$ in $\cal{D}_{\rm T}$ using the knowledge in $\cal{D}_{\rm S}$ and $\cal{T}_{\rm S}$, where $\cal{D}_{\rm S} \ne \cal{D}_{\rm T}$ and/or $\cal{T}_{\rm S} \ne \cal{T}_{\rm T}$.

Typically, the transfer learning can be categorized under three subsettings\cite{surveyTL}: \emph{inductive transfer learning}, \emph{transductive transfer learning} and \emph{unsupervised transfer learning}, based on different situations between the source and target domains and tasks. In particular, the \emph{inductive transfer learning} is adopted in the scenario where $\cal{T}_{\rm S}$ and $\cal{T}_{\rm T}$ are different (regardless whether $\cal{D}_{\rm S}$ and $\cal{D}_{\rm T}$ are the same or not). It is observed that the problem \eqref{p1} and the problem \eqref{p2} are different tasks on the same domain. Therefore, we considers  the inductive transfer learning. There are several approaches to realize the inductive transfer learning, among which we adopts the \emph{parameter-transfer} approach as the two tasks share the domain  and hyperparameters of the proposed model.

\begin{figure}
\centering
\includegraphics[width=0.48\textwidth]{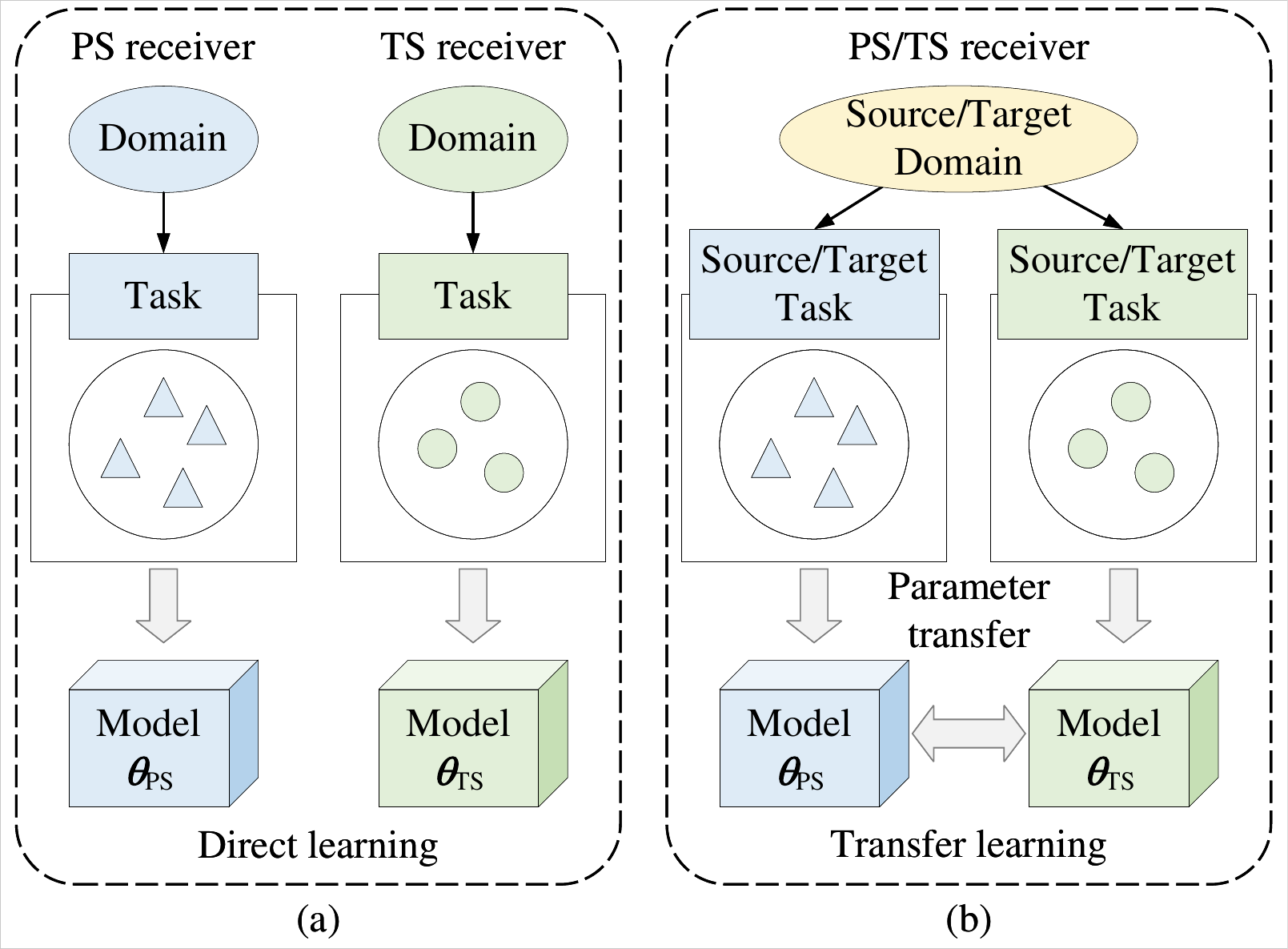}
 \caption{The illustration of (a) direct learning and (b) transfer learning.}\label{TL}
\end{figure}

The parameter-transfer approach is to learn or fine-tune the target parameters ${\bm \theta}_{\rm T}$ with the given source parameters ${\bm \theta}_{\rm S}$,
\begin{flalign}\label{thetaUpdate}
    {\bm \theta}_{\rm T} =\arg \min _{\bm \theta} {\cal L}_{\rm T}\left({\bm \theta}\left|{\bm \theta}_{\rm S}\right.\right)
\end{flalign}
where ${\cal L}_{\rm T}(\cdot)$ represents the target loss function.

Under this transfer learning setting, $\cal{T}_{\rm S}$ and $\cal{T}_{\rm T}$ are correlated in the sense that for optimizing target objective mapping function. And for successful transfer learning between ${\cal T}_{\rm S}$ and ${\cal T}_{\rm T}$, the transfered knowledge need to be general. To mathematically formalize this general concept, we introduce the concept of \emph{common information} \cite{TLCommon}, which we denote by $I({\bm \theta})$, with the following knowledge for ${\cal T}_{\rm S}$ and ${\cal T}_{\rm T}$:
\begin{flalign}
   \{{\bf w}_n, {\rho}_n\} = \Pi_{{\bm \theta}_{\rm S}}\left({\bf h}_n\right), \\
   \{{\bf w}_n, {\alpha}_n\} = \Pi_{{\bm \theta}_{\rm T}}\left({\bf h}_n\right),
\end{flalign}

The \emph{common information} $I({\bm \theta})$ is a function of ${\bm \theta}$ such that it satisfies the following:
\begin{flalign}
   \exists F_1(\cdot) F_2(\cdot) \nonumber \\
   \text{s.t. } {\bm \theta}_{\rm S} &= F_1(I({\bm \theta})),\label{com1} \\
  {\bm \theta}_{\rm T} &= F_2(I({\bm \theta})),  \label{com2}
\end{flalign}
where, $F_1(\cdot), F_2(\cdot)$ denote the neural network mapping, specifically,  $F_1(\cdot)$ represents identity mapping when $I({\bm \theta}) = {\bm \theta}_{\rm S}$, at which point, according to \eqref{thetaUpdate}, we know that $F_2(\cdot)$ must exists to satisfy \eqref{com2}. Thus, we deploy successful transfer learning between ${\cal T}_{\rm S}$, i.e., PS receiver and ${\cal T}_{\rm T}$, i.e., TS receiver.

\subsection{Transfer Learning Between PS and TS Receivers}

This paper intends to transfer knowledge across the task associated with the problem \eqref{p1} (for PS receivers) and the task associated with problem \eqref{p2} (for TS receivers). The goal is to improve the learning performance of the target objective mapping function for the problem \eqref{p1}/\eqref{p2} by transferring knowledge of parameters learned for the problem \eqref{p2}/\eqref{p1}.

In particular, for the domain shared by the two tasks,  $\mathbb{G}^G$ represents the space of all channel vectors, ${\bf g}$ represents a particular learning sample, i.e., $\{{\bf  h}_i\}$, and $P({\bf g})$ represents the channel distribution, e.g., Rayleigh distribution and Rice distribution. For the tasks, the label spaces are not required while the mapping function is represented by the SWIPTNet.

We first train the SWIPTNet to solve the problem \eqref{p1}/\eqref{p2} with learnable parameters ${\bm \theta}_{{\rm PS}}/{\bm \theta}_{\rm TS}$ on the training set. Then, we update ${\bm \theta}_{{\rm PS}}/{\bm \theta}_{\rm TS}$ to solve the problem \eqref{p2}/\eqref{p1}  via the parameter-transfer approach with the following processes.

\subsubsection{Validation and test set} A new validation and test set for the target task is required while the training set keeps unchanged.

\subsubsection{Loss function} Update the loss function from \eqref{l1}/\eqref{l2} to \eqref{l2}/\eqref{l1} to accommodate training for the target task.

\subsubsection{Loading pre-trained model} Use the well-trained ${\bm \theta}_{{\rm PS}}/{\bm \theta}_{\rm TS}$ to initialize the learnable parameters of SWIPTNet for the target task.

\subsubsection{Model training} The parameter-transfer learning can be expressed as
\begin{flalign}\label{theta update}
    {\bm \theta}_{{\rm TS}/{\rm PS}}^{\star} =\arg \min _{\bm\theta}{\cal L}^{\rm TS/PS}\left(\{{{{\bf w}_i\left({\bm \theta}\left|{\bm \theta}_{{\rm PS}/{\rm TS}}\right.\right)}} \}\right)
\end{flalign}
where ${\bm \theta}_{{\rm TS}/{\rm PS}}^{\star}$ denotes the well-trained parameters of the SWIPTNet for the target task. During training phase, the validation set is used to tune the learnable parameters.

The detailed processes of transfer learning is summarized in Algorithm \ref{TS algo}, where, $E$ denotes the training epochs and $B$ denotes the batch sizes.

\begin{algorithm}[t]   \label{TS algo}
\caption{Transfer learning for PS/TS receiver}
{\bf Training dataset:} $\mathcal{R}=\{{\bf V}^{(n)},{\bf E}^{(n)} \}_{n=1}^{N}$\;
{\bf Input:}  $\mathcal{R}$,  ${\bm \theta}_{{\rm PS}/{\rm TS}}$\;
{\bf Output:}  ${\bm \theta}_{{\rm TS}/{\rm PS}}^{\star}$\;
{\bf Initialize}  ${\bm \theta}_{{\rm PS}/{\rm TS}}$\;
 \For{{\rm epoch} $e \in [0,1,\dots,E]$}{
 \For {each batch $b \in [0, 1, ..., B]$} {
 Sample the $b$-th batch $\{{\bf V}^{(n)},{\bf E}^{(n)} \}_{n=1}^{N}$\;
 Obtain beamforming vectors $\{{{{\bf{w}}_n}}| {\bm \theta}_{{\rm PS}/{\rm TS}}\}_{n=1}^{N}$ via the SWIPTNet with ${\bm \theta}_{{\rm PS}/{\rm TS}}$\;
 Calculate the loss function ${\cal L}^{{\rm TS}/{\rm PS}}({\bf{w}}_n)$ via (\ref{l2})\;
 Update ${\bm \theta}_{{\rm TS}/{\rm PS}}^{\star}$ via (\ref{theta update})\;
 }
}
{\bf Return} ${\bm \theta}_{{\rm TS}/{\rm PS}}^{\star}$.
\end{algorithm}

\section{NUMERICAL RESULTS}

\begin{table}[t]
\centering
\begin{threeparttable}\label{stru}
\caption{The structure of the SWIPTNet.}
\begin{tabular}{c|c|c|c|c|c|c}
    \hline
    No. & Type & IFs & OFs & AHs & SELU & BN\\
    \hline
    \hline
    1 & GAL & $N_{\rm T}+N$ & 32 & 20 & $\checkmark$ & $\times$ \\
    2 & GAL & 32$\times$20 & 64 & 20 & $\checkmark$ & $\times$ \\
    3 & FL & 64$\times$20 & 64 & \text{-} & $\checkmark$ & $\times$ \\
    4 & GAL & 64 & 128 & 20 & $\checkmark$ & $\times$ \\
    5 & FL & 128$\times$20 & 128 & \text{-} & $\checkmark$ & $\times$ \\
    6 & GAL & 128 & 256 & 20 & $\checkmark$ & $\times$ \\
    7 & FL & 256$\times$20 & 1024 & \text{-} & $\checkmark$ & $\checkmark$ \\
    8 & FL & 1024 & 512 & \text{-} & $\checkmark$ & $\checkmark$ \\
    9 & FL & 512 & $N_{\rm T}$ & \text{-} & $\times$ & $\times$ \\
    \hline
\end{tabular}
        \begin{tablenotes}[flushleft]
            \footnotesize
            \item IFs/ OFs/ AHs: Input features/ Output features/ Number of attention heads.
        \end{tablenotes}
    \end{threeparttable}
\end{table}

This section provides numerical results to evaluate the proposed DL-based approach as well as the transfer learning. The structure of the SWIPTNet under test is given by Table \ref{stru}.

\subsection{Simulation Setup}
\subsubsection{Simulation scenario and dataset} The number of UEs is set as $N\in\{12,13,14,15\}$. The number of antennas of transmitter is set as $N_{\rm T}=16$.  The power budget of the transmitter is set as $P_{\rm max}=30$ dBm while the individual rate requirement and EH requirement of each UE are set as ${R}_{\rm req}=0.1$ bit/s/Hz and ${{\gamma}_{\rm req}}=-30$ dBm, respectively. The baseband processing noise power and path-loss attenuation at each UE are set as $\sigma_{\rm s}^2=-30$ dBm and $\text{PL}=40$ dB, respectively.

Each training sample includes $N$ complex-value channel vectors, i.e.,  $\{ {\bf h}_{n}\in{\mathbb C}^{N_{\rm T}} \}_{n\in {\cal N}}$. In addition to the channel vector, each validation or test sample includes one label which represents the optimal result of  problem \eqref{p1} or \eqref{p2}. In this paper, we prepare $9$ datasets as shown in Table \ref{Datasets}.
\begin{table}[t]
\centering
\caption{Datasets.}
\label{Datasets}
    \begin{threeparttable}
\begin{tabular}{c|c|c|c|c|c}
\hline
No. &$N_{\rm T}$ &$N$ & Receiver & Size& Type\\
 \hline
  \hline
 1   &16&12& PS/TS & 100,000& A\\
 \hline
 2   &16&12& PS &6,000& B\\
 \hline
 3   &16&13& PS &6,000& B\\
 \hline
 4   &16&14& PS & 6,000& B\\
 \hline
 5   &16&15& PS & 6,000& B\\
 \hline
 6   &16&12& TS & 6,000& B\\
 \hline
 7   &16&13& TS & 6,000& B\\
 \hline
 8   &16&14& TS & 6,000& B\\
 \hline
 9   &16&15& TS & 6,000& B\\
 \hline
\end{tabular}
 \begin{tablenotes}
        \footnotesize
        \item {Type A: Only training set is included.}
        \item {Type B: Only test and validation set is included and the splitting ratio is $0.33:0.66$.}
\end{tablenotes}
    \end{threeparttable}
\end{table}

\subsubsection{Computer configuration} All DL models  are trained and tested by Python 3.8 with Pytorch 1.10.1 on a computer with Intel(R) Xeon(R) Plstinum 8255C CPU and NVIDIA RTX 2080 Ti (11 GB of memory).

\subsubsection{Initialization and training} The learnable weights are initialized according to Xavier (Glorot) method \cite{xavier} and the learning rate is initialized as $10^{-4}$. The $Adam$ method \cite{Adam} is adopted as the optimizer during the training phase. The batch size is set as $16$ for $50$ epochs of training. The learnable weights with the best performance are used as the training results.

\subsubsection{Baseline}
In order to evaluate the SWIPTNet numerically, the following four baselines are considered, i.e.,
\begin{itemize}
\item \textbf{CVX-based approach:}  An iterative CVX optimization algorithm, similar to \cite{bk3}.
\item \textbf{MLP:}  A basic feed-forward  neural network by replacing the GAL in the SWIPTNet by FL, similar to \cite{MLP}.
\item \textbf{GCN:}  A basic GNN by replacing the GAL in the SWIPTNet by graph convolutional operation, similar to \cite{GNN2021}.
\item \textbf{GAT:}  A basic GCN with multi-head attention mechanism, similar to \cite{GAT2024}.
\end{itemize}

\subsubsection{Performance metrics}  The following metrics are considered to evaluate the DL models on the test sets.
\begin{itemize}
  \item [1)]
  \emph{Optimality performance:} The ratio of the average achievable sum-rate (with feasible solutions) by the DL model to the optimal sum-rates with $N_{\rm Te} = N_{\rm Tr}$ where $N_{\rm Te}$ and $N_{\rm Tr}$ represent the numbers of UEs in test set and training set, respectively.
  \item [2)]
  \emph{Scalability performance:} The ratio of the average achievable sum-rate (with feasible solutions) by the DL model to the optimal sum-rates with $N_{\rm Te} > N_{\rm Tr}$.
  \item [3)]
  \emph{{Feasibility rate}:} The percentage of the feasible solutions to the considered problem by the DL model.
  \item [4)]
  \emph{Inference time:} The average running time required to calculate the feasible solution with the given channel vectors by the DL model.
\end{itemize}

\subsection{Ablation Experiment}

\begin{table*}[ht]
    \centering
    \caption{Ablation experiment for PS receiver: $N_{\rm Tr}=12$.}
    \label{AblationPS}
    \begin{threeparttable}
    \begin{tabular}{ccc|c|c|c|c|c|c|c|c}
 \hline
        \multirow{3}{*}{LM} & \multirow{3}{*}{SO} & \multirow{3}{*}{LC} & \multicolumn{8}{c}{$N_{\rm Te}$}\\
        \cline{4-11}
        & & &\multicolumn{2}{c|}{12} & \multicolumn{2}{c|}{13} & \multicolumn{2}{c|}{14} & \multicolumn{2}{c}{15} \\
        \cline{4-11}
        & & & OP & FR & SC & FR & SC & FR & SC & FR \\
        \hline
        \hline
        $\times$ & $\checkmark$ & $\checkmark$ & 82.97\% & 99.20\% & 76.97\% &89.88\% & 75.63\% & 85.13\% & 71.69\% & 78.10\% \\
        \cline{4-11}
        $\checkmark$ & $\times$ & $\checkmark$ & 91.96\% & 99.14\% & 89.92\% &98.14\% & 88.35\% & 97.13\% & 87.27\% & 94.16\% \\
        \cline{4-11}
        $\checkmark$ & $\checkmark$ & $\times$ & 85.16\% & 99.29\% & 83.36\% &95.01\% & 82.03\% & 89.80\% & 78.95\% & 87.92\% \\
        \cline{4-11}
        $\checkmark$ & $\checkmark$ & $\checkmark$ & {\bf 93.27\%} & 99.48\% & {\bf 91.03\%} &99.34\% & {\bf 90.00\%} & 99.32\% & {\bf 88.79\%} & 98.17\% \\
        \hline
    \end{tabular}
    \begin{tablenotes}
        \footnotesize
        \item {LT/SO/SL: Laplace transform/ Single-type output/ layer connection.}
        \item {OP/SC/FR: Optimality performance/ scalability performance/ feasibility rate.}
    \end{tablenotes}
    \end{threeparttable}
\end{table*}

\begin{table*}[ht]
    \centering
    \caption{Ablation experiment for TS receiver: $N_{\rm Tr}=12$.}
    \label{AblationTS}
    \begin{tabular}{ccc|c|c|c|c|c|c|c|c}
        \hline
        \multirow{3}{*}{LM} & \multirow{3}{*}{SO} & \multirow{3}{*}{LC} & \multicolumn{8}{c}{$N_{\rm Te}$}\\
        \cline{4-11}
        & & &\multicolumn{2}{c|}{12} & \multicolumn{2}{c|}{13} & \multicolumn{2}{c|}{14} & \multicolumn{2}{c}{15} \\
        \cline{4-11}
        & & & OP & FR & SC & FR & SC & FR & SC & FR \\
        \hline
        \hline
        $\times$ & $\checkmark$ & $\checkmark$ & 84.12\% & 99.17\% & 81.98\% &99.24\% & 81.49\% & 92.16\% & 75.51\% & 86.11\% \\
        \cline{4-11}
        $\checkmark$ & $\times$ & $\checkmark$ & 93.01\% & 99.50\% & 90.98\% &99.21\% & 89.90\% & 96.54\% & 87.92\% & 91.36\% \\
        \cline{4-11}
        $\checkmark$ & $\checkmark$ & $\times$ & 86.19\% & 99.50\% & 84.85\% &93.73\% & 83.90\% & 91.10\% & 79.54\% & 88.64\% \\
        \cline{4-11}
        $\checkmark$ & $\checkmark$ & $\checkmark$ & {\bf 93.79\%} & 99.50\% & {\bf 91.98\%} &99.15\% & {\bf 90.73\%} & 98.89\% & {\bf 88.16\%} & 97.17\% \\
        \hline
    \end{tabular}
\end{table*}

Table \ref{AblationPS} and Table \ref{AblationTS} give the ablation results to show the effectiveness of the Laplace transform \eqref{eh_f}, single-type output \eqref{cons:gamma} or \eqref{dlalpha} and layer connection (Table \ref{stru}) in the proposed model for PS and TS receivers, respectively. It is observed that the Laplace transform helps to improve the learning performance as it can be regarded as a parameter-free message passing to enable each node to learn the interaction with its neighboring nodes (cf. Remark \ref{Rem2}). Besides, although the single-type  output have little impact on the optimality performance and scalability performance, it enhances the feasibility rate. The reason is that by \eqref{cons:gamma}/\eqref{dlalpha}, the output ports for PS/TS ratios is relaxed and the EH constraint \eqref{primal:B}/\eqref{p2:B} is replaced by \eqref{cons:gamma} which tends to be non-trivial to be satisfied (cf. Remark \ref{Rem0}). Moreover, by stacking the GAL and FL alternatively, the learning performance is also enhanced since the FL between two GALs not only reduces the input dimension of the following GAL but also mitigates over-smoothing issue (cf. Remark \ref{remark3}). At last, by comparing Table \ref{AblationPS} and Table \ref{AblationTS}, one can see that the learning performance  for TS receiver slightly outperforms that for PS receiver. The reason could be that the computation of achievable information rate  is independent of TS ratio in \eqref{p6} but deeply coupled with PS ratio in \eqref{rps}.

\subsection{Generalization Performance}

Table \ref{test PS} and Table \ref{test TS} evaluate the generalization performance of the SWIPTNet for the PS and TS receivers, respectively. Particularly, the generalization performance stands for the optimality and the scalability performance during the test phase. For comparison, the results of the four baselines are also given.

For the optimality performance, the proposed SWIPTNet achieves the best optimality performance with the highest feasibility rate compared with other DL models for both the PS and TS receivers, which demonstrates the effectiveness of the proposed designs. It is observed that the SWIPTNet and GAT are obviously superior to the MLP and the GCN as they adopt attention-enabled aggregation to explore the hidden interaction due to the inter-UE interference. Besides, the optimality performance gap between the SWIPTNet and the CVX-based approach is less than $8\%$ for the PS receiver and $7\%$ for the TS receiver with almost no loss of the feasibility rate. However, the inference time of the DL-based models are nearly a thousand times faster than the CVX-based approach. That is, the SWIPTNet is able to achieve near-optimal and real-time inference at the cost of pre-training. Fortunately, the training of the SWIPTNet can be implemented offline like other DL applications.

For the scalability performance, the SWIPTNet outperforms other DL models for both the PS and TS receivers with a steady feasibility rate. However, the well-trained MLP fails to be scalable to the number of UEs due to its fixed input ports. Particularly, the SWIPTNet trained on the dataset with $N_{\rm Tr} = 12$ (Dataset No. 1) can also achieve a good performance within $7\%$ scalability performance loss for the PS receiver and $6\%$ performance loss for the TS receiver. Besides, the max scalability performance loss of GCN and GAT are  $10\%$ and $11\%$ for the PS receiver, $11\%$ and $10\%$ for the TS receiver, respectively when tested on the dataset with $N_{\rm Te} \in \{13, 14, 15\}$ (Dataset No. 3, 4, 5 and 7, 8, 9). Note that such a scale capability is of high importance due to the dynamic nature of the wireless network. Moreover, it is observed that both the scalability performance and feasibility performance are degraded with the increase of UEs for the GNN-based models, and the performance loss goes larger as with the increase of $|N_{\rm Te}-N_{\rm Tr}|$. Nevertheless, the SWIPTNet shows the good robustness to unseen problem sizes.

\begin{table*}[t]
    \centering
    \caption{Generalization performance evaluation for PS receiver: $N_{\rm Tr}=12$ and $ N_{\rm Te}\in\{12,13,14,15\}$.}
    \label{test PS}
    \begin{threeparttable}
    \begin{tabular}{c|cc|cc|cc|cc|cc}
\hline
      \multirow{2}{*}{$N_{\rm Te}$} &  \multicolumn{2}{c|}{CVX} &  \multicolumn{2}{c|}{MLP}  & \multicolumn{2}{c|}{GCN}  & \multicolumn{2}{c|}{GAT} & \multicolumn{2}{c}{SWIPTNet} \\
    \cline{2-11}
       &  \multicolumn{1}{c|}{OP/SC} & FR & \multicolumn{1}{c|}{OP/SC} & FR & \multicolumn{1}{c|}{OP/SC} & FR & \multicolumn{1}{c|}{OP/SC}  & FR & \multicolumn{1}{c|}{OP/SC}  & FR \\
\hline
\hline
     12  & \multicolumn{1}{c|}{{100\%}$^{\dag}$} & \multicolumn{1}{c|}{100\%} & \multicolumn{1}{c|}{{25.67\%}$^{\dag}$} & \multicolumn{1}{c|}{100\%} & \multicolumn{1}{c|}{{65.16\%}$^{\dag}$} & \multicolumn{1}{c|}{99.13\%} & \multicolumn{1}{c|}{{82.97\%}$^{\dag}$} &
     \multicolumn{1}{c|}{99.20\%} &
     \multicolumn{1}{c|}{{\bf 93.27\%}$^{\dag}$} & \multicolumn{1}{c}{99.48\%} \\
    \hline
     13  & \multicolumn{1}{c|}{100\%} & \multicolumn{1}{c|}{100\%} & \multicolumn{1}{c|}{$\times$} & \multicolumn{1}{c|}{$\times$} & \multicolumn{1}{c|}{61.67\%} & \multicolumn{1}{c|}{94.27\%} & \multicolumn{1}{c|}{76.97\%} &
     \multicolumn{1}{c|}{89.88\%} &
     \multicolumn{1}{c|}{\bf 91.03\%} & \multicolumn{1}{c}{99.34\%} \\
     \hline
     14  & \multicolumn{1}{c|}{100\%} & \multicolumn{1}{c|}{100\%} & \multicolumn{1}{c|}{$\times$} & \multicolumn{1}{c|}{$\times$} & \multicolumn{1}{c|}{59.04\%} & \multicolumn{1}{c|}{88.79\%} & \multicolumn{1}{c|}{75.63\%} &
     \multicolumn{1}{c|}{85.13\%} &
     \multicolumn{1}{c|}{\bf 90.00\%} & \multicolumn{1}{c}{99.32\%} \\
    \hline
     15 & \multicolumn{1}{c|}{100\%} & \multicolumn{1}{c|}{100\%} & \multicolumn{1}{c|}{$\times$} & \multicolumn{1}{c|}{$\times$} & \multicolumn{1}{c|}{55.73\%} & \multicolumn{1}{c|}{85.92\%} & \multicolumn{1}{c|}{71.69\%} &
     \multicolumn{1}{c|}{78.10\%} &
     \multicolumn{1}{c|}{\bf 88.79\%} & \multicolumn{1}{c}{98.17\%} \\
    \hline
    \hline
     IT &  \multicolumn{2}{c|}{2.164s} &  \multicolumn{2}{c|}{3.185ms}  & \multicolumn{2}{c|}{3.315ms}  & \multicolumn{2}{c|}{3.857ms} & \multicolumn{2}{c}{3.671ms}\\
     \hline
\end{tabular}
 \begin{tablenotes}
        \footnotesize
            \item {$^{\dag}$: The result marked with $^{\dag}$ represents the optimality performance.}
            \item {IT: Inference time.}
    \end{tablenotes}
\end{threeparttable}
\end{table*}

\begin{table*}[t]
    \centering
    \caption{Generalization performance evaluation for TS receiver: $N_{\rm Tr}=12$ and $ N_{\rm Te}\in\{12,13,14,15\}$.}
    \label{test TS}
    \begin{tabular}{c|cc|cc|cc|cc|cc}
\hline
      \multirow{2}{*}{$N_{\rm Te}$} &  \multicolumn{2}{c|}{CVX} &  \multicolumn{2}{c|}{MLP}  & \multicolumn{2}{c|}{GCN}  & \multicolumn{2}{c|}{GAT} & \multicolumn{2}{c}{SWIPTNet} \\
    \cline{2-11}
       &  \multicolumn{1}{c|}{OP/SC} & FR & \multicolumn{1}{c|}{OP/SC} & FR & \multicolumn{1}{c|}{OP/SC} & FR & \multicolumn{1}{c|}{OP/SC}  & FR & \multicolumn{1}{c|}{OP/SC}  & FR \\
\hline
\hline
     12  & \multicolumn{1}{c|}{{100\%}$^{\dag}$} & \multicolumn{1}{c|}{100\%} & \multicolumn{1}{c|}{{27.84\%}$^{\dag}$} & \multicolumn{1}{c|}{99.49\%} & \multicolumn{1}{c|}{{67.58\%}$^{\dag}$} & \multicolumn{1}{c|}{98.19\%} & \multicolumn{1}{c|}{{84.12\%}$^{\dag}$} &
     \multicolumn{1}{c|}{99.17\%} &
     \multicolumn{1}{c|}{{\bf 93.79\%}$^{\dag}$} & \multicolumn{1}{c}{99.50\%} \\
    \hline
     13  & \multicolumn{1}{c|}{100\%} & \multicolumn{1}{c|}{100\%} & \multicolumn{1}{c|}{$\times$} & \multicolumn{1}{c|}{$\times$} & \multicolumn{1}{c|}{64.37\%} & \multicolumn{1}{c|}{93.31\%} & \multicolumn{1}{c|}{81.98\%} &
     \multicolumn{1}{c|}{99.24\%} &
     \multicolumn{1}{c|}{\bf 91.98\%} & \multicolumn{1}{c}{99.15\%} \\
     \hline
     14  & \multicolumn{1}{c|}{100\%} & \multicolumn{1}{c|}{100\%} & \multicolumn{1}{c|}{$\times$} & \multicolumn{1}{c|}{$\times$} & \multicolumn{1}{c|}{62.17\%} & \multicolumn{1}{c|}{88.62\%} & \multicolumn{1}{c|}{81.49\%} &
     \multicolumn{1}{c|}{92.16\%} &
     \multicolumn{1}{c|}{\bf 90.73\%} & \multicolumn{1}{c}{98.89\%} \\
    \hline
     15 & \multicolumn{1}{c|}{100\%} & \multicolumn{1}{c|}{100\%} & \multicolumn{1}{c|}{$\times$} & \multicolumn{1}{c|}{$\times$} & \multicolumn{1}{c|}{59.31\%} & \multicolumn{1}{c|}{86.76\%} & \multicolumn{1}{c|}{75.51\%} &
     \multicolumn{1}{c|}{86.11\%} &
     \multicolumn{1}{c|}{\bf 88.16\%} & \multicolumn{1}{c}{97.17\%} \\
    \hline
     IT &  \multicolumn{2}{c|}{3.306s} &  \multicolumn{2}{c|}{1.514ms}  & \multicolumn{2}{c|}{2.633ms}  & \multicolumn{2}{c|}{3.033ms} & \multicolumn{2}{c}{2.962ms}\\
     \hline
\end{tabular}
\end{table*}

\subsection{Transfer Learning}

\begin{figure}[t]
\centering
\includegraphics[width=0.48\textwidth]{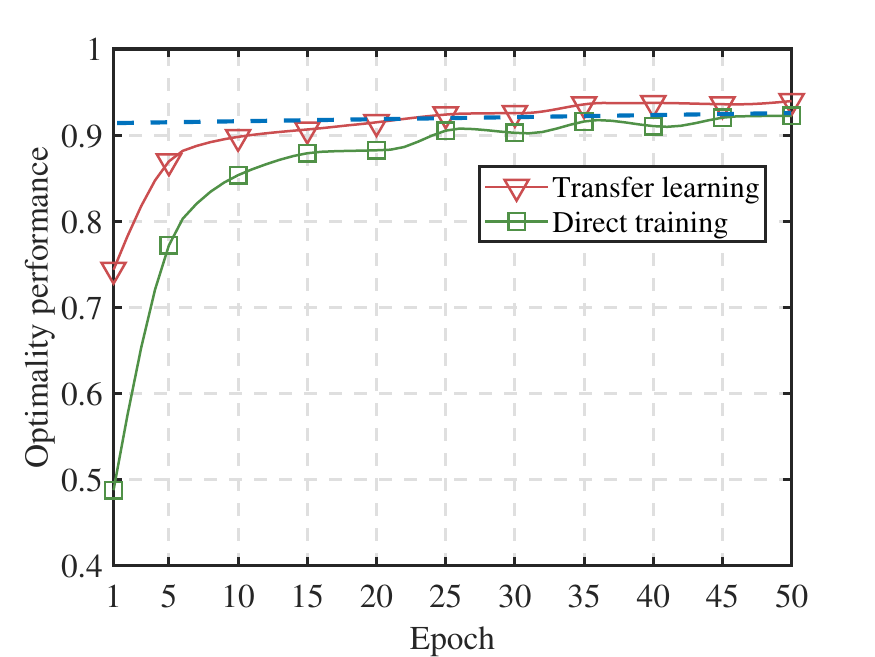}
 \caption{Convergence behavior of the transfer learning and direct training.}\label{TL_TS_Result}
\end{figure}

Fig. \ref{TL_TS_Result} shows the convergence behavior of transfer learning  and direct training. For the transfer learning, we use the well-trained parameters of the SWIPTNet for the PS receiver as the initial parameters and fine-tune the parameters for the TS receiver on the Dataset No. 1. For direct training, we train the SWIPTNet on the Dataset No. 1 with randomly initialized parameters. It is observed that the transfer learning after $20$ epochs is able to achieve similar optimality performance to the direct training after $50$ epochs. That is, the convergence speed of the transfer learning is much faster than that of the direct training. After $50$ epochs, a performance gain is obtained by the transfer learning compared with the direct learning. The reason can be that the similarity between the PS and TS receivers allows knowledge transferring via parameters, and the learning performance can be improved thanks to being trained from parameters with better initial performance\eqref{theta update}.

\begin{table}[t]
    \centering
    \caption{The performance of the transfer learning and direct training: $N_{\rm Tr}=12$ and $ N_{\rm Te}=12$.}
    \label{TL test}
    \begin{tabular}{c|c|c|c}
    \hline
    \multicolumn{2}{c|}{Approach} & OP & FR\\
    \hline
    \multicolumn{2}{c|}{Direct training} & 93.79\% & 99.50\%\\
    \hline
    \multirow{3}{*}{Transfer learning} & 10 epochs & 89.75\% & 99.49\% \\
    \cline{2-4}
    & 30 epochs & 93.86\% & 99.44\% \\
    \cline{2-4}
    & 100 epochs & 94.80\% & 99.31\% \\
    \hline
\end{tabular}
\end{table}

Table \ref{TL test} gives the optimality performance of transfer learning and direct training. The results of the SWIPNet via transfer learning after $10$, $30$ and $100$ epochs are given. It is observed that the transfer learning after $10$ epochs is inferior to the direct training while the transfer learning after $30$ epochs already outperforms the direct training, and the performance gain increases with the epochs. The reason is that the two considered tasks share the training set, and epochs of transfer learning can be roughly regarded as the sum of the epochs of training for both source and target tasks. Together with Fig. \ref{TL_TS_Result}, the transfer learning is able to improve the learning performance with less training cost.

\section{Conclusion}

This paper proposes a unified DL framework for SWIPT, which is applicable to both the PS and TS receivers. A GNN-based model termed SWIPTNet has been designed and trained via unsupervised learning to map the CSI vectors to beamforming vectors and PS/TS ratios to maximize the QoS-constrained sum-rate.
Transfer learning have been further adopted to transfer knowledge between the PS and TS receivers. The ablation experiment has validated the effectiveness of the Laplace transform, single-type output and feature transformation involved in the SWIPTNet, and thus, the SWIPTNet outperformed existing models. Moreover, the SWIPTNet has been shown to not only achieve near-optimal performance with millisecond-level inference speed but also be scalable to the number of UEs. Finally, the transfer learning has demonstrated to help the SWIPTNet to accelerate training convergence and improve the generalization performance.


\begin{thebibliography}{99}

\bibitem{bk1}
K. W. Choi et al., ``Simultaneous wireless information and power transfer (SWIPT) for Internet of Things: Novel receiver design and experimental validation," \emph{IEEE Internet Things J.}, vol. 7, no. 4, pp. 2996-3012, Apr 2020.


\bibitem{bk2}
R. Zhang and C. K. Ho, ``MIMO broadcasting for simultaneous wireless information and power transfer," \emph{IEEE Trans. Wireless Commun.}, vol. 12, no. 5, pp. 1989-2001, May 2013.

\bibitem{bk3}
Y. Lu, K. Xiong, P. Fan, Z. Zhong, B. Ai, and K. B. Letaief, ``Worst-case energy efficiency in secure SWIPT networks with rate-splitting ID and power-splitting EH receivers," \emph{IEEE Trans. Wireless Commun.}, vol. 21, no. 3, pp. 1870-1885, Mar 2022.


\bibitem{bk5}
Y. Lu, K. Xiong, P. Fan, Z. Zhong, and K. B. Letaief, ``Coordinated beamforming with artificial noise for secure SWIPT under non-linear EH model: Centralized and distributed designs," \emph{IEEE J. Sel. Areas Commun.}, vol. 36, no. 7, pp. 1544-1563, Jul 2018.


\bibitem{dl1}
Y. Lu, W. Mao, H. Du, O. A. Dobre, D. Niyato, and Z. Ding, ``Semantic-aware vision-assisted integrated sensing and communication: Architecture and resource allocation," \emph{IEEE Wireless Commun.},  vol. 31, no. 3, pp. 302-308, Jun 2024.

\bibitem{dl2}
Y. Li, Y. Lu, R. Zhang, B. Ai, and Z. Zhong, ``Deep learning for energy efficient beamforming in MU-MISO networks: A GAT-based approach," \emph{IEEE Wireless Commun. Lett.}, vol. 12, no. 7, pp. 1264-1268, Jul 2023.

\bibitem{dl3}
Y. Li, Y. Lu, B. Ai, Z. Zhong, D. Niyato, and Z. Ding, ``GNN-enabled max-min fair beamforming," \emph{IEEE Trans. Veh. Technol.}, vol. 73, no. 8, pp. 12184-12188, Aug 2024.

\bibitem{MLP}
H. Sun, X. Chen, Q. Shi, M. Hong, X. Fu, and N. D. Sidiropoulos, ``Learning to optimize: Training deep neural networks for interference management,” \emph{IEEE Trans. Signal Process.}, vol. 66, no. 20, pp. 5438-5453, Oct 2018.

\bibitem{CNN}
Q. Hu, Y. Cai, Q. Shi, K. Xu, G. Yu, and Z. Ding, ``Iterative algorithm induced deep-unfolding neural networks: Precoding design for multiuser MIMO systems," \emph{IEEE Trans. Wireless Commun.}, vol. 20, no. 2, pp. 1394-1410, Feb 2021.

\bibitem{unsuper}
T. Lin and Y. Zhu, ``Beamforming design for large-scale antenna arrays using deep learning," \emph{IEEE Wireless Commun. Lett}, vol. 9, no. 1, pp. 103-107, Jan 2020.

\bibitem{unsuper2}
C. Sun, C. She, and C. Yang. ``Unsupervised deep learning for optimizing wireless systems with instantaneous and statistic constraints," arXiv.2006.01641, 2020.

\bibitem{lugnn}
Y. Lu et al., ``Graph neural networks for wireless networks: Graph representation, architecture and evaluation," early accessed in
 \emph{IEEE Wireless Commun.}, 2024.

\bibitem{GNN2021}
Y. Shen, Y. Shi, J. Zhang, and K. B. Letaief, ``Graph neural networks for scalable radio resource management: Architecture design and theoretical analysis," \emph{IEEE J. Sel. Areas Commun.}, vol. 39, no. 1, pp. 101-115, Jan 2021.

\bibitem{GCN2023}
Y. Shen, J. Zhang, S. H. Song, and K. B. Letaief, ``Graph neural networks for wireless communications: From theory to practice," \emph{IEEE Trans. Wireless Commun.}, vol. 22, no. 5, pp. 3554-3569, May 2023.

\bibitem{GNN2024}
K. Liang, G. Zheng, Z. Li, K. -K. Wong, and C. -B. Chae, ``A data and model-driven deep learning approach to robust downlink beamforming optimization," early accessed in \emph{IEEE J. Sel. Areas Commun.}, 2024.

\bibitem{GAT2024}
Y. Li, Y. Lu, B. Ai, O. A. Dobre, Z. Ding, and D. Niyato, ``GNN-based beamforming for sum-rate maximization in MU-MISO networks," \emph{IEEE Trans. Wireless Commun.}, vol. 23, no. 8, pp. 9251-9264, Aug 2024.


\bibitem{icnet}
C. He, Y. Li, Y. Lu, B. Ai, Z. Ding, and D. Niyato, ``ICNet: GNN-enabled beamforming for MISO interference channels with statistical CSI," \emph{IEEE Trans. Veh. Technol.}, vol. 73, no. 8, pp. 12225-12230, Aug 2024.


\bibitem{hg}
Z. Song, Y. Lu, X. Chen, B. Ai, Z. Zhong, and D. Niyato, ``A deep learning framework for physical-layer secure beamforming," early accessed in \emph{IEEE Trans. Veh. Technol.}, 2024.


\bibitem{DNNSWIPT}
T. -H. Vu, T. -V. Nguyen, and S. Kim, ``Cooperative NOMA-enabled SWIPT IoT networks with imperfect SIC: Performance analysis and deep learning evaluation," \emph{IEEE Internet Things J.}, vol. 9, no. 3, pp. 2253-2266, Feb 2022.

\bibitem{DNN2SWIPT}
H. T. Thien, P. -V. Tuan, and I. Koo, ``Deep learning-based secure transmission for SWIPT system with power-splitting scheme," in Proc. \emph{ICTC}, pp. 50-55, Dec 2021.

\bibitem{DNNS}
M. R. Camana, C. E. Garcia, and I. Koo, ``Deep learning-assisted power minimization in underlay MISO-SWIPT systems based on rate-splitting multiple access," \emph{IEEE Access}, vol. 10, pp. 62137-62156, Jun 2022.

\bibitem{DNNSWIPT2}
W. Lee, K. Lee, H. -H. Choi, and V. C. M. Leung, ``Deep learning for SWIPT: Optimization of transmit-harvest-respond in wireless-powered interference channel," \emph{IEEE Trans. Wireless Commun.}, vol. 20, no. 8, pp. 5018-5033, Aug 2021.

\bibitem{TLwithLoss}
W. Cui and W. Yu, ``Transfer learning with reconstruction loss," \emph{IEEE Trans. Mach. Learn. Commun. Networking.}, vol. 2, pp. 407-423, Apr 2024.

\bibitem{CSI}
J.-M. Kang, C.-J. Chun, I.-M. Kim, and D. I. Kim, ``Deep RNN based channel tracking for wireless energy transfer system,” \emph{IEEE Syst. J.}, vol. 14, no. 3, pp. 4340-4343, Sep 2020.

\bibitem{non-linear}
Y. Lu, K. Xiong, P. Fan, Z. Zhong, and K. B. Letaief, ``Robust transmit beamforming with artificial redundant signals for secure SWIPT system under non-linear EH model," \emph{IEEE Trans. Wireless Commun.}, vol. 17, no. 4, pp. 2218-2232, Apr 2018.

\bibitem{over-smoothing}
M. Liu, H. Gao, and S. Ji, ``Towards deeper graph neural networks," in Proc. \emph{ACM SIGKDD}, pp. 338-348, 2020.

\bibitem{surveyTL}
S. J. Pan and Q. Yang, ``A survey on transfer learning," \emph{IEEE Trans. Knowl. Data Eng.}, vol. 22, no. 10, pp. 1345-1359, Oct 2010.

\bibitem{TLCommon}
W. Cui and W. Yu, ``Transfer learning with input reconstruction Loss," in Proc. \emph{IEEE Globecom}, pp. 615-620, Jan 2022.

\bibitem{xavier}
X. Glorot and Y. Bengio, ``Understanding the difficulty of training deep feedforward neural networks" in Proc. \emph{JMLR}, pp. 249-256, Jan 2010.

\bibitem{Adam}
D. P. Kingma and J. Ba, ``Adam: A method for stochastic optimization,” in Proc. \emph{ICLR}, pp. 1–15, Feb 2015.







\end{thebibliography}
\end{document}